\documentclass[lettersize,journal]{IEEEtran}
\usepackage{amsmath,amsfonts}
\usepackage{algorithmic}
\usepackage{algorithm}
\usepackage{array}
\usepackage{flushend}
\usepackage[caption=false,font=normalsize,labelfont=sf,textfont=sf]{subfig}
\usepackage{textcomp}
\usepackage{stfloats}
\usepackage{multirow}
\usepackage{url}
\usepackage{verbatim}
\usepackage{graphicx}
\usepackage{cite}
\usepackage{xcolor}
\usepackage{pifont}
\usepackage{soul}
\usepackage{arydshln}
\newcommand{\cmark}{\ding{51}}%
\newcommand{\xmark}{\ding{55}}%
\begin{document}

\title{CVSM: Contrastive Vocal Similarity Modeling}

\author{Christos Garoufis, Athanasia Zlatintsi, and Petros Maragos

\thanks{Christos Garoufis is with the School of Electrical and Computer Engineering, National Technical University of Athens, Athens, Greece and with the Robotics Institute, Athena Research Center, Athens, Greece (e-mail: cgaroufis@mail.ntua.gr).
Athanasia Zlatintsi is with the Robotics Institute, Athena Research Center, Athens, Greece (e-mail: nancy.zlatintsi@athenarc.gr).
Petros Maragos is with the School of Electrical and Computer Engineering, National Technical University of Athens, Athens, Greece, the Robotics Institute, Athena Research Center, Athens, Greece and HERON - Hellenic Robotics Center of Excellence, Athens, Greece (e-mail: maragos@cs.ntua.gr).}}



\maketitle

\begin{abstract}
The availability of large, unlabeled datasets across various domains has contributed to the development of a plethora of methods that learn representations for multiple target (downstream) tasks through self-supervised pre-training. In this work, we introduce CVSM (Contrastive Vocal Similarity Modeling), a contrastive self-supervised procedure for music signal representation learning in the audio domain that can be utilized for musical and vocal similarity modeling. Our method operates under a contrastive framework, maximizing the similarity between vocal excerpts and musical mixtures containing the same vocals; we devise both a \textit{label-informed} protocol, leveraging artist identity information to sample the contrastive pairs, and a \textit{label-agnostic} scheme, involving artificial mixture creation from randomly sampled vocal and accompaniment excerpts, which are paired with vocals from the same audio segment.
We evaluate our proposed method in measuring vocal similarity both objectively, through linear probing on a suite of appropriate downstream tasks, and subjectively, via conducting a user study consisting of pairwise comparisons between \textcolor{black}{different models in a recommendation-by-query setting}. Our results indicate that the representations learned through CVSM are effective in musical and vocal similarity modeling, outperforming numerous baselines across both isolated vocals and complete musical mixtures. Moreover, while the availability of artist identity labels during pre-training leads to overall more consistent performance both in the evaluated downstream tasks and the user study, a label-agnostic CVSM variant incorporating hybrid pre-training with real and artificial mixtures achieves comparable performance to the label-informed one in artist identification and perceived vocal similarity.
\end{abstract}


\section{Introduction}

Throughout all history, from the musical practices of the ancient world and classical orchestral music to the heavily produced music of our era, musical pieces have always constituted multi-faceted expressions of art. Multiple performers, each using a different instrument or their voices, cooperate in order to create compound sounds, usually in harmonic and rhythmic consonance. However, each of the co-playing sources exhibits its own distinct characteristics, influencing the musical piece in a specific way, such as defining the rhythm of the piece, exhibiting the technical proficiency of the performers, or attempting to elicit a certain mood. Thus, successfully modeling the similarity between particular musical sources is of profound importance, with applications ranging from music analysis~\cite{fu10} to music recommendation systems~\cite{deldjoo24}.

Possibly the most expressive ``instrument'' in a musical piece, thanks to the range of different mental states and emotions it can convey, is the human voice~\cite{juslin03}. Perceptually, vocal similarity can be traced to numerous high-level attributes, including the phonation, the articulation, the timbral clarity as well as the vocal prosody~\cite{sundberg77,obin16}. These high-level perceptual traits can be connected partially to low-level acoustic features that translate well into timbral similarity, including the fundamental frequency (F0), formant frequencies (F1-F4), their normalized amplitude ratios, as well as measures related to the harmonic regularity, volume, and spectral noise~\cite{perrachione19,lee19,kreiman21,liu24}. Thus, early attempts to model vocal similarity~\cite{obin16} have focused on extracting, or annotating, particular descriptive features from vocal clips, and training machine learning classifiers to directly predict them, creating thus an acoustic feature space that could be utilized for vocal sample retrieval. More recently, it has been shown that such feature spaces can be obtained by neural networks that have been directly trained from speech, with a suitable end-to-end objective~\cite{liu24}. 

However, modeling the singing voice entails a number of challenges compared to spontaneous speech, including higher frequency range and harmonics amplitude, as well as higher rhythmicality and vowel duration~\cite{mesaros10,zhang21,yakura22}. This, coupled with the presence of instrumental accompaniment, has steered early attempts in modeling vocals in the presence of background music into necessitating either suitable feature selection, or application of a source separation pre-processing step~\cite{fujihara10, humphrey18}. Another challenge regards the subjectivity of systematically evaluating whether different voices are similar or dissimilar. As a result, apart from subjective listening tests~\cite{fujihara10}, vocal similarity has been usually evaluated through other proxy tasks, such as singer identification~\cite{fujihara10,yakura22,torres23,yamamoto23}, vocal technique recognition~\cite{yakura22, yamamoto23}, gender recognition~\cite{yakura22} or vocal pitch estimation~\cite{yamamoto23}.

\IEEEpubidadjcol

Recently, the availability of large, unlabeled datasets across various domains and the upsurge of deep learning, coupled with the upscaling of hardware resources, have given rise to the field of self-supervised learning (SSL). Under SSL frameworks, neural networks are trained to learn representations, which discriminate between different input samples without having access to their target labels. SimCLR~\cite{simclr19}, which can be viewed as a batch-wide variant of triplet metric learning~\cite{hoffer15}, constitutes one of the most prominent SSL paradigms. In short, SimCLR-based methods aim to create a latent space where projections of different views of the same sample, typically generated by devising a suitable augmentation pipeline, are close to each other, while simultaneously being distant from projections of semantically different samples. Whilst originally developed for computer vision applications, SimCLR has been transferred to other domains~\cite{saeed21,mohsenvand20} or even multimodal settings~\cite{radford21,elizalde23} by appropriate choices of the backbone encoder and the applied augmentations.

In the domain of musical audio, a number of research works have applied contrastive learning in order to model vocal similarity~\cite{yakura22,torres23} using clean vocal excerpts. However, the contrastive pre-training pipelines followed do not employ instrumental accompaniment, hindering generalization in the case of commercial music where background instruments are also present. Potential detours to this problem would involve coupling song excerpts with isolated vocals during the pre-training stage~\cite{garoufis23}, leading to a latent space that successfully correlates attributes of the isolated vocals with the musical piece, or using jointly clean and mixture pairs~\cite{desblancs24} during pre-training, reducing the domain gap between musical mixtures and vocal excerpts. Nonetheless, neither of the above strategies guarantees invariance to instrumental accompaniment by disentangling the vocal properties in the learned latent space. 

In this work, we present CVSM\footnote{To foster reproducibility of our experimental results, we make our source code as well as pre-trained model weights available at: https://github.com/cgaroufis/CVSM} 
(Contrastive Vocal Similarity Modeling), a framework that can be utilized for vocal retrieval and vocal similarity modeling both in clean vocals and in-the-wild (i.e., in the presence of background instrumental accompaniment). The presented framework relies on pairing vocal excerpts with mixtures of vocal excerpts and instrumental accompaniment. It is based on MSCOL~\cite{garoufis23}, but improves on it in terms of robustness, specificity and generalizability. In short, our main contributions are the following:
\begin{itemize}
\item We leverage of the availability of artist identity labels, as their utilization has proven effective in general-purpose music representation learning~\cite{alonso22,park17}, by proposing a \textit{label-informed} pre-training scheme. In this case, contrastive pairs are created by matching excerpts of isolated vocals with full song excerpts, originating from the same artist\footnote{We note that since artist labels are utilized to guide the contrastive sampling, the proposed label-informed scheme is not strictly self-supervised.}, deviating from~\cite{alonso22} where both elements of the pair are sampled from the complete musical mixture. 
\item Inspired by data augmentation techniques used in the field of music source separation~\cite{uhlich17,kong21}, we also design a \textit{label-agnostic} pre-training protocol, wherein we generate artificial musical mixtures as anchors by superimposing the positive vocal sample into an accompaniment from a different, randomly sampled musical piece. This way, we intend to create a latent space invariant to instrumental properties of the musical piece~\cite{mccallum24}; this approach is similar to~\cite{lee19k}, but applies itself in a fully self-supervised, contrastive setting. In order to alleviate the domain gap between the artificial mixtures used for pre-training and original music pieces, we also experiment with i)~stochastic application of the proposed augmentation, selecting randomly (on-the-fly) either artificial or real musical mixtures during pre-training and ii) contrastive self-supervised fine-tuning, using solely pairs of real musical mixtures and the corresponding vocals.
\end{itemize}

The framework was pre-trained using the publicly available Music4All~\cite{santana20} dataset, and was evaluated both i) objectively, by training shallow downstream classifiers upon the learned embeddings in the tasks of gender identification and artist identification and similarity, and ii) subjectively, by evaluating the learned latent space of the framework in retrieving and recommending musical pieces with vocals similar to a given query through a listening test. Our findings indicate that CVSM can effectively model musical and vocal similarity across both isolated vocals and complete musical mixtures. In particular, the \textit{label-informed} CVSM variant outperforms its respective baseline~\cite{alonso22} in isolated vocals, while performing comparably to it in the presence of instrumental accompaniment. We also show that even without the availability of artist identity labels during pre-training, CVSM can create a latent space that effectively conveys artist identity information, outperforming label-agnostic baselines~\cite{saeed21,garoufis23,li23} in artist identification and artist similarity. In fact, when pre-trained using a combination of real and artificial mixtures, the performance of the \textit{label-agnostic} CVSM variant approaches the one achieved through label-informed pre-training in artist identification. 
These results are further corroborated by the conducted user study, with the CVSM variant pre-trained with artist-guided sampling scoring the highest among all evaluated models in modeling both overall and vocal similarity in musical mixtures. Interestingly, despite its middling performance in overall perceived similarity, label-agnostic pre-training incorporating a hybrid strategy of utilizing both real and artificial musical mixtures throughout its training pipeline performs comparably to label-informed models in encapsulating perceived vocal similarity, suggesting that it can be utilized for vocal-based retrieval applications without prior access to artist metadata. 

\section{Related Work}

Traditionally, vocal similarity in musical pieces has been modeled via tags denoting the presence or absence of vocals, or inherent attributes of the vocalist~\cite{kim20}. Hence, acoustic features and, more recently, time-frequency representations have been extracted from mixed audio excerpts and fed to shallow machine learning classifiers, or deep neural networks, to directly predict those attributes~\cite{marques11,choi16}. Thanks to recent developments in the task of music source separation~\cite{kong21,htdemucs,schulze23}, which have been fueled by the release of diverse datasets~\cite{musdb18,li18,pereira23}, isolating the vocals from the mixed audio~\cite{sharma19,deberardinis20,garoufis24} has emerged as a promising alternative, leading to more robust estimation of vocal tags~\cite{sharma19,garoufis24}. However, these methods introduce computational overhead, through the deployment of an auxiliary network to estimate the vocals from the mixed audio.

Recent advances in SSL have led to the creation of large-scale music foundation models~\cite{li23,won24}, trained to learn general-purpose representations of audio through large amounts of unlabeled data; contrastive learning, in particular, has been identified as a promising avenue towards this goal. The majority of contrastive approaches follow the SimCLR~\cite{simclr19} scheme of projecting batches of paired input views into a shared latent space, which has been shown to outperform other self-supervised approaches~\cite{niizumi21, anton23} in music tagging tasks~\cite{brocal24}. COLA~\cite{saeed21} has set a simple, yet effective paradigm for transferring SimCLR into the audio domain, making use of a data sampling strategy consisting of cropping different excerpts from the same audio sample. As the encoder backbone,~\cite{saeed21} adapts EfficientNet-B0~\cite{tan19}, initially developed for image classification, to audio understanding tasks by accepting spectrogram inputs.

The aforementioned data sampling strategy usually forms the basis of more complex augmentation chains. These chains often include additional augmentations such as gain amplification, frequency masking/filtering, reverberation effects, time warping or pitch shifting operations applied either on time-frequency input representations~\cite{xiao22,mccallum24,choi22}, or on the waveform itself~\cite{spijkervet21,vasquez22}. The pair creation process may also be assisted by auxiliary supervision in the form of pseudo-labels; these have included editorial metadata, such as artist or album information~\cite{alonso22,meehan25} and playlist co-occurrence statistics~\cite{alonso23,meehan25}. Furthermore, numerous of the above methods deviate from~\cite{saeed21} in the choice of encoder backbone; while the Efficient-Net encoder is indeed a popular choice~\cite{alonso22},~\cite{brocal24} experimented with a Res-Net~\cite{he16} backbone, while~\cite{xiao22} opted for a SWin-Transformer~\cite{liu21}.

Learning the identity of artists, either using complete musical pieces~\cite{park17,alonso22} or isolated singing voices~\cite{yakura22,torres23}, has been recently employed as an indirect way to model musical or vocal similarity~\cite{fujihara10,humphrey18}. Such approaches rely either on supervised learning, where neural networks correlate audio excerpts to artist identity labels~\cite{park17}, or on contrastive learning, building upon the SimCLR framework by pairing elements from the same audio segment~\cite{yakura22, torres23} or artist~\cite{alonso22}. However, the majority of those operate either in the space of singing voices~\cite{yakura22,torres23}, which hinders their ability to model the vocals in the presence of background music, or in complete musical mixtures~\cite{alonso22,park17}, being thus unable to disentangle attributes pertaining to the singing voice. A potential solution to this could involve the utilization of both complete musical mixtures and isolated sources during the training process~\cite{lee19k,desblancs24}. Given that particular attributes of musical pieces, such as the tempo, the primary melody, or the target elicited emotion, are tied to either specific instruments, or the vocals~\cite{deberardinis20,garoufis23,garoufis24}, a number of frameworks have been developed, which associate, through a contrastive process, segments of musical pieces with isolated sources. For instance, information related to the tempo can be captured through percussive components~\cite{heydari22,desblancs22}, whereas associating the mixture segments with randomly chosen isolated sources can lead in effective general-purpose representation learning of musical signals~\cite{garoufis23}. Moreover, encoders pre-trained in musical source association have been employed for evaluating the plausibility of automatically generated instrumental accompaniments in~\cite{ciranni24}. 

However, the majority of those methods operate using either raw, or weakly augmented, views of the input audio segments; a less explored family of augmentations, inspired by mixup~\cite{zhang16}, involves creating additive mixtures of audio signals as pre-training inputs. Its most basic variant, involving generation of synthetic audio excerpts via direct superimposition of distinct audio segments, has been applied in various works~\cite{srivastava22,mccallum22,niizumi21} as a general-purpose input augmentation. Yet, despite the wide usage, and success, of generating artificial mixtures by superimposing source excerpts of different origins in tasks such as frame-wise pitch estimation~\cite{gfeller20,riou23} and music source separation~\cite{samuel20,kong21,htdemucs}, this avenue has only concurrently been explored in contrastive setups~\cite{cheston25}, for the task of music sample identification. Moreover, artificial increase of the training dataset by generating synthetic mixtures has been reported to introduce a domain gap during inference~\cite{tremblay18, tobin17}; as such, researchers have attempted to bypass it by either performing some small-scale finetuning on datasets from the target domain~\cite{simon22} or introducing an alignment stage, with regard to either tempo, or pitch, before superimposing the various audio segments~\cite{zhang21,kratimenos20}.

\section{Proposed Method}

\begin{figure*}[t]

        \centering        \centerline{\includegraphics[width=17cm]{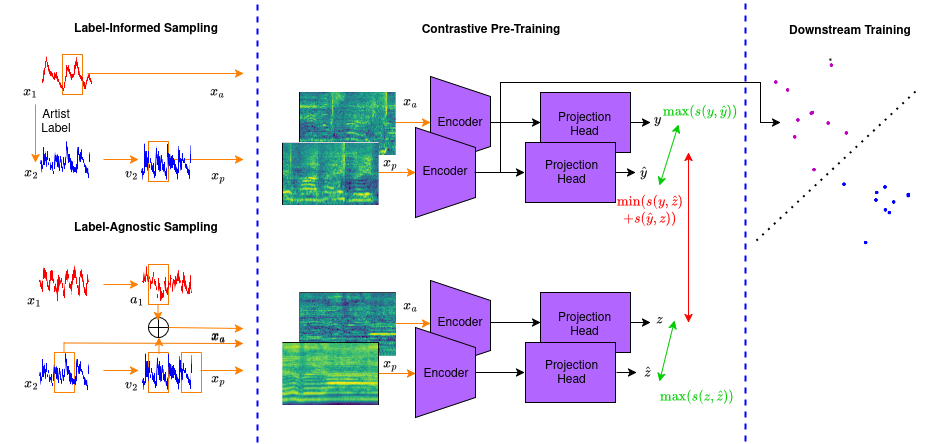}}
    \vspace{-0.25cm}
    \caption{\textcolor{black}{Overview of our proposed framework for learning audio representations. Contrastive pairs are generated either using (top left) label-informed sampling, where pairs of musical mixtures (consisting of vocals and instrumental accompaniment) and isolated vocals are sampled from the same artist, or (bottom left) in a label-agnostic manner, by i) creating artificial song mixtures by superimposing the vocals and accompaniments of different song excerpts or b) sampling excerpts from the complete song, and coupling them with time-shifted excerpts of the vocals. These contrastive pairs are then used to pre-train an encoder backbone with a contrastive loss objective (right). }}
    \label{fig:ovrl}
    \vspace{-0.5cm}
\end{figure*}

In this work, we propose CVSM, an extension of MSCOL~\cite{garoufis23} for vocal similarity modeling; an overview of the proposed framework is presented in Fig.~\ref{fig:ovrl}. Both MSCOL and CVSM were designed for learning representations of musical audio by associating audio excerpts with isolated sources of the audio through a batch-wise contrastive loss objective. However, while MSCOL was developed for general-purpose music representation learning, here we focus solely on the case of vocals. To this end, we extend MSCOL by modifying the contrastive pair generation process as well as the followed training scheme, both by incorporating artificial mixture generation~\cite{uhlich17} and employing artist-level pair sampling~\cite{alonso22,desblancs24}, in order to increase the robustness of the framework and its invariance to non-vocal elements.

\subsection{Contrastive Pair Generation}

\textbf{Label-Informed Sampling}: The approach we follow for sampling batches of contrastive pairs is built upon the segment-wise sampling procedure followed by COLA~\cite{saeed21}. In more detail, we couple complete musical mixtures with isolated vocal excerpts\footnote{Throughout the paper, we use the terms \textit{excerpts} for network input audio, \textit{segments} for the audio slices used to crop the audio excerpts, and \textit{previews/clips} for complete audio files.}, with the mixture and vocal excerpts originating from the same artist. Employing artist identity labels to guide the pair selection process helps in projecting audio segments with similar vocals close to each other; moreover, in contrast to segment-level sampling, the learned similarity is not tied to song-specific attributes, such as the rhythm or the key. No augmentations are applied in this stage, so as to capture both timbre-related and pitch-related information about the singing voice~\cite{yakura22}. 

\textbf{Label-Agnostic Sampling}: The availability of artist labels guides sampling towards pairs which, while containing vocals from the same artist, do disentangle vocal attributes from non-vocal information, since they may originate from different songs. However, when following the procedure outlined above in a label-agnostic setting, the mixtures and isolated vocals are highly correlated in terms of various non-vocal properties, since they are sampled from the same segment. Thus, in this case the contrastive pairs are created by generating artificial mixtures, consisting of a vocal excerpt superimposed with a randomly sampled instrumental accompaniment, and coupling them with isolated vocals from the same audio segment. Using randomly selected accompaniments to generate artificial mixtures has shown to increase the robustness of networks in frame-wise singing voice understanding tasks~\cite{gfeller20,riou23}. 
Furthermore, compared to pairing the vocal sample with its original accompaniment~\cite{garoufis23}, this strategy i) not only acts as \textit{data augmentation}, increasing the network's ability to generalize~\cite{uhlich17,kong21}, but also ii) helps in forming a latent space \textit{invariant to non-vocal elements} of musical pieces~\cite{mccallum24}, being thus more suitable for vocal modeling. Similar to above, no further augmentations are applied to either the isolated vocals or the artificial mixtures.

\subsection{Network Backbone and Projection Head} 

As the encoder, we make use of the EfficientNet~\cite{tan19} family of models. Its combination of small parameter footprint and solid learning capability renders it a suitable choice for contrastive learning pipelines~\cite{saeed21,garoufis23,ciranni24,alonso22,torres23}, which are dependent on larger batch sizes~\cite{simclr19,saeed21,spijkervet21}. Since EfficientNets consist of 2D convolutions, the input waveforms have to be transformed into the time-frequency domain; thus, after the input waveforms are generated, their mel-spectrograms are computed, with 64 bands, a window length of 25~ms and a hop size of 10~ms, before being fed to the network. 

EfficientNet architectures consist of distinct blocks (stages), each of which processes its input through a series of depthwise-separable convolutions~\cite{chollet17} incorporating inverted residual connections and downsamples it through a two-dimensional pooling operation. The output tensor of the final convolutional stage is flattened through a global average pooling operation, in order to acquire a temporally and spectrally invariant feature vector, which can be used in conjunction with a classification head for downstream tasks. In our case, we make use of the EfficientNet-B0 encoder, which amounts to a total of 1.5M parameters, including 9 stages and 18 convolutional layers, and leading to an 1280-dimensional embedding. 

For the projection head, we follow~\cite{saeed21}; thus, we apply a linear layer, with a dimensionality of 512, on top of the encoder, followed by Layer Normalization~\cite{ba16} and a tanh() activation. We note that this projection head is used only during pre-training, for the purposes of measuring the similarity between semantically similar embeddings, and discarded during downstream training.

\subsection{Training Scheme and Loss Function}

For pre-training, we generate contrastive pairs, following the procedure outlined in Sec.~III-A, and train the network in identifying the vocal excerpt that is included in each mixture (either real or artificial) in the respective batch. For this purpose, first the bilinear similarity, $s(y,\hat{y})$~\cite{saeed21}, is computed between all anchor and positive embeddings in each batch. These similarities are first transformed into logits, via a softmax operation, and then used to train the projection head, using the
normalized binary cross-entropy loss for all elements in each batch $S$:

\begin{equation}
\mathcal{L} = -\sum_{y \in S}{\mathrm{log} \dfrac{\mathrm{exp}( s(y,\hat{y}))}{\sum_{z\in S} (\mathrm{exp}(s(y,z)))}},
\end{equation}
where $s(y,\hat{y}) = y^T W\hat{y}$ denotes the aforementioned bilinear similarity between the anchor embeddings $y$ and the positive embeddings $\hat{y}$ computed through a learnable linear layer $W$. We note that, in contrast to MSCOL~\cite{garoufis23}, since we are mostly interested in identifying timbral differences between vocal excerpts, rather than acquiring information about high-level attributes through the existence or absence of vocals, we do not use the modified cross-entropy loss presented in~\cite{garoufis23}.

Whereas in the \textit{label-informed} case the same training protocol is applied throughout the whole pre-training duration, we empirically noted that, for the \textit{label-agnostic} pre-training, the learned embeddings did not generalize well in practice, especially for shorter audio clips. We hypothesize that the core reason for this is the domain gap~\cite{simon22} that incurs between the artificial data used for pre-training and the real data used in practice. To alleviate the gap, we experiment with the following strategies:

\begin{itemize}
\item \textbf{Hybrid Pre-Training:} In this case, we simultaneously expose the network to real and artificial mixtures during pre-training. To this end, real and artificial mixtures for each vocal anchor are generated stochastically, at probabilities $p$ and $1-p$, so that each batch of contrastive pairs contains both real and artificial musical mixtures.

\item \textbf{In-Domain Finetuning}: Here, we
introduce a second training stage in the self-supervised training procedure, after pre-training with artificial mixtures. During this stage, we no longer use artificial mixtures as anchor examples, feeding instead the network solely with pairs of song excerpts with vocal excerpts isolated from the same segment.
\end{itemize}

\section{Experimental Setup}

\subsection{Data and Preprocessing}

As our primary dataset, we employed Music4All~\cite{santana20}, a publicly available dataset consisting of metadata (such as artist names and song titles), lyrics, genre information and user listening statistics for a large-scale music catalog, as well as 30~sec audio previews (clips) of the included songs. In total, the dataset includes 109,269 songs from 16,269 different artists, at diverse sampling rates. 

\begin{figure}[t!]
\begin{minipage}{0.49\linewidth}
        \centering
        \centerline{\includegraphics[width=4.5cm]{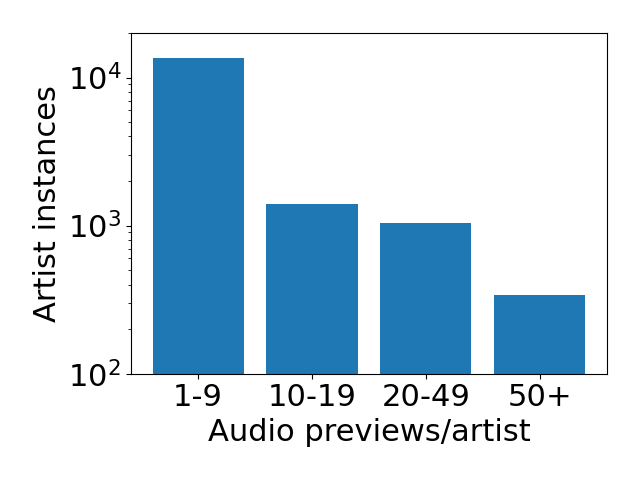}}
    \end{minipage}
    \begin{minipage}{0.49\linewidth}
        \centering
        \centerline{\includegraphics[width=4.5cm]{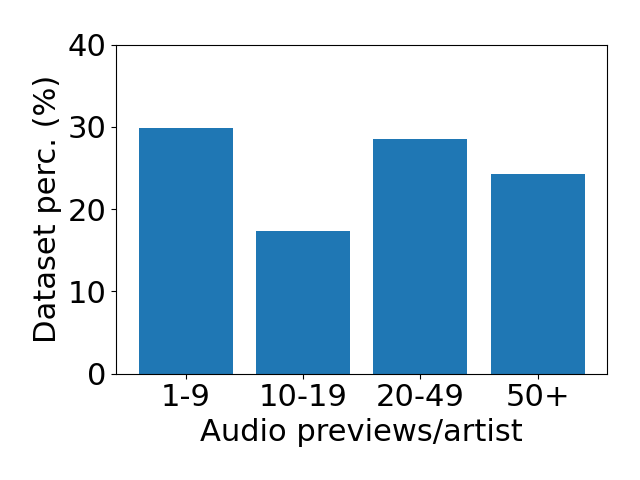}}
\end{minipage}
\vspace{-0.25cm}
    \caption{Dataset statistics for Music4All: the number of artist identities (left) and the percentage of audio previews in the dataset (right), grouped according to the number of audio previews available for each artist.}
    \vspace{-0.65cm}
    \label{fig:dataspread}
\end{figure}

In Fig.~~\ref{fig:dataspread}, we depict the exact distribution of the artist identities in the dataset according to the number of audio previews available for each artist (left), as well as the percentage of the available previews corresponding to each category (right). From the left subfigure, we observe that the distribution of artist labels in Music4All is not balanced; only 2.08 \% of the artists are represented with more than 50 audio previews each, whereas {82.92 \%} of the artists present in the dataset have less than 10 previews each. The subfigure on the right further supports this point, since comparable portions of the dataset were sampled from artists with less than 10 previews (32,598 total previews, or 29.83\%) and artists with more than 50 previews (26,507 previews total, or 24.26\%). 

Similar to other works utilizing large datasets of singing voices for music understanding tasks~\cite{garoufis23,yakura22}, an open-source framework for music source separation, open-unmix~\cite{stoter19}, was chosen for the extraction of the vocal segments and the instrumental accompaniment, facilitating both the contrastive pre-training objective and the creation of the artificial mixtures. All audio clips, prior to pre-training, were downsampled to 16~kHz, for computational efficiency purposes, and split into 5~sec segments. Vocal segments with a mean amplitude lower than 0.01 (amounting to 20.53\% of the full dataset) were discarded.

\subsection{Training and Evaluation Protocols}

We pre-train CVSM embeddings with Music4All, following the protocol delineated in Sec.~III. During pre-training, the dataset is split into training, validation and testing subsets, using an 8:1:1 ratio, so that there is no artist leakage between the different subsets. The backbone encoder was pre-trained for 8,000 steps (approx. 160 epochs for the subset of Music4All containing vocals), with each step consisting of 64 mini-batches of 128 contrastive pairs. For the in-domain finetuned backbone, we interrupt pre-training with artificial mixtures at 6,000 steps, and only use real mixture-vocal pairs for the final 2,000 steps. We used Adam~\cite{Kingma} as the optimizer, with an initial learning rate of 0.001. The pre-training progress was monitored by measuring the loss of the pre-text task in intervals of 10 steps; the learning rate was halved in case the running average of the validation loss did not improve over 1,000 steps.

In order to evaluate the capability of the proposed framework to model vocal similarity, we freeze the model's encoder and benchmark the performance of its learned embeddings at the following vocal understanding tasks, as per the concurrent literature~\cite{torres23,yakura22,yamamoto23}:
\begin{itemize}
 \item \textit{Gender Identification}: The goal in this task is to correctly classify the biological gender (male/female) of the singing artist. For this task, we re-use Music4All,  \textcolor{black}{using only artists for whom gender information is retrievable following~\cite{ferraro21}}. For evaluation purposes, we employ an artist-stratified 10-fold cross-validation procedure, so as to gauge the timbral content of the embeddings on unseen data. As our metric, we employ the classification accuracy (Acc., \%). 

 \item \textit{Artist Identification}: Similar to the previous case, the goal is to correctly classify the artist identity given an embedding vector. We utilize the testing subset of Music4All, using the available previews from $M=50$ randomly sampled artists from the testing subset (for which no label information has been available during pre-training), and performing 5~repetitions of the experiment with differently sampled artists. During each repetition, an 8:1:1 split of the respective dataset into training, validation and testing data is applied. In contrast to the case of gender identification, the task target regards discrimination between audio excerpts of artists that have been accessed during downstream training; thus, we opt for a random data split, instead of an artist-stratified one. Since the distribution of artist labels in Music4All is not balanced, we report on both the identification accuracy as well as the macro-F1 score, which can be calculated as the average of the per-class F1 scores.

  \item \textit{Artist Similarity}: In this case, we directly probe the learned embeddings, computing the similarity between pairs that belong to the same, or different, artists. Similar to the previous case, we utilize the testing subset of Music4All for this task. Following~\cite{torres23}, we compute the Equal Error Rate (EER) and Mean Normalized Rank (MNR) metrics (as defined in~\cite{torres23,yang21}), which are used for retrieval purposes~\cite{torres23,lattner22} and denote the ability of a system to identify input sample pairs as similar (from the same origin) or dissimilar (from different origins), penalizing low similarities between samples from the same origin. For EER calculation, we use $K=5000$ sets of similar and dissimilar pairs; for MNR, we use a batch size of $N=50$, and $K=100$ trials.
 
\end{itemize}

We note that the network backbone is always kept frozen, and the performance is measured through training a linear classifier upon the learned embeddings (or, in the case of artist similarity, directly probing the learned latent space). All experiments were repeated for two different configurations; i) for the complete musical pieces (in the presence of instrumental accompaniment), using all 1~sec excerpts with vocals present, as well as ii) on isolated vocals.
For the gender and artist identification tasks, network performance is measured over each 30~sec clip, by aggregating the estimates for each excerpt that contains vocals into a single prediction. The downstream linear classifiers are trained over a maximum of 200 epochs, using again Adam~\cite{Kingma}, a learning rate of 5e-4, and early stopping by monitoring the clip-wise identification accuracy, with a patience of 6 epochs. For the artist similarity task, we use the cosine similarity as the similarity function, measured between pairs of averaged embeddings of all valid (containing vocals) 1~sec excerpts within each clip; the averaged embeddings are L2-normalized prior to the distance calculation.

\subsection{Baselines}

We compare CVSM to three identity-agnostic baselines, which learn excerpt-level audio representations:
\begin{itemize}
\item COLA~\cite{saeed21}, trained using time-shifted pairs of song excerpts.
\item MSCOL~\cite{garoufis23}, trained to associate song excerpts with the corresponding vocals in a single-source setup. 
\item MERT~\cite{li23}, trained via a masked language modelling (MLM) self-supervised setup, with target pseudo-labels provided by a combination of acoustic and musical teacher models.
\end{itemize}

As well as the identity-informed baseline:
\begin{itemize}
\item COLA-ART, where we follow the sampling procedure used in~\cite{alonso22}, associating song excerpts with the same artist label. 
\end{itemize}
For COLA and MSCOL, as well as COLA-ART, we trained the baseline encoders using the same pre-training dataset and under the same protocol as CVSM; for MERT, we make use of the publicly available checkpoint obtained via pre-training in Music4All\footnote{https://huggingface.co/m-a-p/MERT-v0-public}, and use embeddings obtained via 1 sec audio excerpts to ensure a fair comparison. 

Regarding CVSM, we experiment with the following variants:

\begin{itemize}

    \item CVSM-A, incorporating label-agnostic pre-training with excerpt-level creation of \textit{artificial} mixtures of vocals and instrumental accompaniment, without exposing the model to any real musical mixtures. 
    \item CVSM-AH, where the network is pre-trained with the \textit{hybrid} scheme of viewing both real and artificial musical mixtures during pre-training. After experimentation, the artificial pair creation probability was set to $p = 0.5$. 
    \item CVSM-AF, where we \textit{finetune} CVSM-A in-domain using solely real mixture-vocal pairs.
    \item CVSM-ART, which has been pre-trained utilizing artist identity information, as delineated in Sec. III-A.
\end{itemize}

An overview of various properties of these models is presented in Tab.~\ref{tab:props}.

\begin{table}[t]
    \begin{center}
                \caption{Overview of the compared methods regarding the inclusion of vocals (first column), artist identity information (second column), and artificial mixtures (third column) in the pre-training pipeline; the fourth column denotes whether artificially pre-trained methods underwent finetuning with only real pairs. - in each cell denotes non-applicability of the respective property for the corresponding method; all methods are self-supervised.}
            \vspace{-0.15cm}
            \renewcommand{\arraystretch}{1.3}
    \begin{tabular}{c||c|c|c|c} 
    Method & Vocals & Artist ID & Artif. Mix & Finetune \\  \hline

    COLA~\cite{saeed21} & \xmark & \xmark & - & - \\ 
    MSCOL~\cite{garoufis23} & \cmark & \xmark & \xmark & - \\ 
    MERT~\cite{li23} & \xmark & \xmark & - & - \\ \hdashline[3pt/1.5pt]
    COLA-ART~\cite{alonso22} & \xmark & \cmark & - & - \\
    \hline
    CVSM-A & \cmark & \xmark & \cmark & \xmark \\ 
    CVSM-AH  & \cmark & \xmark & \cmark & \xmark \\ 
    CVSM-AF  & \cmark & \xmark & \cmark & \cmark \\ \hdashline[3pt/1.5pt]
    CVSM-ART & \cmark & \cmark & \xmark & -  
    \label{tab:props}
    \end{tabular}
    \end{center}
    \vspace{-0.8cm}
\end{table} 

\subsection{Listening Test}

To further validate the results obtained by the above evaluation, we also assessed the ability of CVSM to model vocal attributes through a subjective listening test. In more detail, the latent space of the networks was probed in order to retrieve, by means of cosine similarity, the most similar musical piece to a given query. Then, participants were presented with pairs of retrieved musical pieces from different networks, along with the given query, and were tasked with responding to the following questions:\begin{itemize} 
\item \textit{Overall Similarity}: Which of the two retrieved musical pieces is more similar to the initial query in terms of overall musicality (encompassing timbral similarity, rhythmic similarity, and general feeling)?  
\item \textit{Vocal Similarity}: Which of the two retrieved pieces is more similar to the initial query regarding the vocals? 
\end{itemize}
  
In total, 37 people took part in the survey, recruited through our social circles, work environments, and community mailing lists, and were informed about the survey's purpose and procedure prior to their participation; no data were recorded apart from the anonymized demographic information and the questionnaire responses. The participants (23 male, 13 female, 1 other) had an average age of 30.92 years ($\pm$ 6.10 years), and were generally familiar with artificial intelligence and its applications (4.16 $\pm$ 0.94) in a 5-point Likert scale. Despite this familiarity and the overall positive relationship of the participants with music (81.08\% of the participants responded to be listening to music in a daily basis), their familiarity with music recommender systems in particular was highly variant (3.30 $\pm$ 1.27 in a 5-point Likert scale).  

During the listening test, pairwise comparisons were conducted between two randomly selected models. The model pool consisted of all models presented in. Tab.~I, with the exception of CVSM-AF as we will discuss afterwards. Each participant was presented with a total of $N=20$ different triplets (10 of complete musical mixtures and 10 of isolated vocals), sampled randomly, for each participant, out of an initial pool of $M=500$ queries, each of a 15~sec duration. Instances where the same song by the selected models was recommended were mostly excluded\footnote{Since the probability of two models retrieving the same recommendation for a given query is not uniform across all models, not all inter-model comparisons were conducted with the same frequency.}, with the exception of a few cases which were left in the listening test as controls.

\section{Objective Evaluation}

\begingroup

\begin{table*}[t]
    \begin{center}
                \caption{Experimental results on the tasks of gender identification, artist identification and artist similarity on Music4All, using complete musical mixtures as input; the first three rows correspond to COLA, MSCOL, and MERT respectively, while row 4 corresponds to the COLA-ART label-informed baseline; rows 5-7 correspond to identity-agnostic CVSM variants, and the final row corresponds to CVSM incorporating artist identity information.}
            \vspace{-0.15cm}
 \renewcommand{\arraystretch}{1.3}
 \begin{tabular}{c||c|c|c|c|c}

    \multirow{2}{*}{Configuration} & \textcolor{black}{Gender ID} & \multicolumn{2}{c|}{\textcolor{black}{Artist ID}} & \multicolumn{2}{c}{\textcolor{black}{Artist Sim.}} \\
    \cline{2-6} &   Acc. (\%) $\uparrow$ &  Acc. (\%) $\uparrow$ & macro-F1 (\%) $\uparrow$  & EER $\downarrow$ & MNR $\downarrow$   \\ \hline 
    COLA~\cite{saeed21} & 81.01 $\pm$ 3.68 & 59.67 $\pm$ 5.02  & 47.02 $\pm$ 3.68 & 29.28 $\pm$ 4.22 & 19.85 $\pm$ 3.39   \\ 
    MSCOL~\cite{garoufis23} & 85.24 $\pm$ 3.80 &70.40 $\pm$ 3.71 & 59.32 $\pm$ 8.35 & 26.66 $\pm$ 4.46 & 17.47 $\pm$ 3.08  \\  
    MERT~\cite{li23} & 81.61 $\pm$ 2.84 & 65.29 $\pm$ 5.65 & 55.85 $\pm$ 7.50 & 35.34 $\pm$ 3.98 & 25.97 $\pm$ 3.44 \\ \hdashline[3pt/1.5pt] COLA-ART~\cite{alonso22} & \textbf{87.15 $\pm$ 3.44} & 76.58 $\pm$ 1.92 & 69.31 $\pm$ 3.43 & 20.24 $\pm$ 3.98 & 9.83 $\pm$ 2.17 \\ \hline 
   CVSM-A  & \textbf{85.65 $\pm$ 3.31} & 73.26 $\pm$ 4.14 & 59.40 $\pm$ 7.05 & 32.10 $\pm$ 3.88 & 20.32 $\pm$ 3.34  \\ 
          CVSM-AH  & 85.40 $\pm$ 3.33 & \textbf{77.66 $\pm$ 2.19} & \textbf{66.30 $\pm$ 4.77} & \textbf{23.96 $\pm$ 3.95} & \textbf{14.02 $\pm$ 2.61}  \\  
    CVSM-AF  & 85.48 $\pm$ 3.12 & 72.79 $\pm$ 2.60 & 60.27 $\pm$ 3.76 & 24.82 $\pm$ 4.08 & 15.33 $\pm$ 2.79  \\ \hdashline[3pt/1.5pt]
    CVSM-ART  & 87.12 $\pm$ 3.41 & \textbf{78.65 $\pm$ 2.24} & \textbf{70.00 $\pm$ 3.30} & \textbf{19.62 $\pm$ 4.07}  & \textbf{9.70 $ \pm$ 2.07}  
    \label{tab:mixres}
    \end{tabular}
    \end{center}
    \vspace{-0.4cm}
\end{table*} 
\endgroup
\begingroup
\begin{table*}[t]
   \renewcommand{\arraystretch}{1.3}
    \begin{center}
                \caption{Experimental results on the tasks of gender identification, artist identification and artist similarity on Music4All, using isolated vocal excerpts as input; the first three rows correspond to COLA, MSCOL, and MERT respectively, while row 4 corresponds to the COLA-ART label-informed baseline; rows 5-7 correspond to identity-agnostic CVSM variants, and the final row corresponds to CVSM incorporating artist identity information.}
            \vspace{-0.15cm}
     \begin{tabular}{c||c|c|c|c|c} 
    \multirow{2}{*}{Configuration} & \textcolor{black}{Gender ID} & \multicolumn{2}{c|} {\textcolor{black}{Artist ID}} & \multicolumn{2}{c}{\textcolor{black}{Artist Sim.}} \\
    \cline{2-6} &   Acc. (\%) $\uparrow$ &  Acc. (\%) $\uparrow$ & macro-F1 (\%) $\uparrow$  & EER $\downarrow$ & MNR $\downarrow$   \\ \hline 
   COLA~\cite{saeed21} & 83.89 $\pm$ 3.38 & 54.85 $\pm$ 4.66 & 40.75 $\pm$ 4.83 & 29.04 $\pm$ 4.85 & 19.33 $\pm$ 2.96  \\ 
   MSCOL~\cite{garoufis23} & \textbf{85.27 $\pm$ 2.87} & 70.67 $\pm$ 4.55 & 56.58 $\pm$ 4.23 & 25.66 $\pm$ 4.44& 15.88 $\pm$ 2.69  \\ 
       MERT~\cite{li23} & 81.67 $\pm$ 1.91 & 62.88 $\pm$ 7.02 & 47.78 $\pm$ 9.50 & 35.28 $\pm$ 3.53 & 27.06 $\pm$ 3.89 \\ \hdashline[3pt/1.5pt]
       COLA-ART~\cite{alonso22} & 85.05 $\pm$ 2.83 & 71.06 $\pm$ 2.38 & 57.45 $\pm$ 3.30 &25.12 $\pm$ 4.28 & 15.74 $\pm$ 2.81  \\ \hline 
    CVSM-A  & 85.23 $\pm$ 3.26 & \textbf{78.20 $\pm$ 3.74} & 65.10 $\pm$ 4.00 & \textbf{23.14 $\pm$ 4.03} & \textbf{13.65 $\pm$ 2.58}  \\ 
          CVSM-AH & \textbf{86.06 $\pm$ 3.20} & 76.33 $\pm$ 3.64 & 66.09 $\pm$ 3.57  & 23.18 $\pm$ 3.75 & 13.71 $\pm$ 2.36  \\ 
           CVSM-AF &84.91 $\pm$ 3.61 &  77.67 $\pm$ 3.30 & \textbf{66.67 $\pm$ 5.46} & 23.86 $\pm$ 4.14 & 14.41 $\pm$ 2.50  \\ \hdashline[3pt/1.5pt] 

    CVSM-ART  & \textbf{87.46 $\pm$ 3.13} & \textbf{78.57 $\pm$ 3.17} & \textbf{68.52 $\pm$ 3.92} & \textbf{19.02 $\pm$ 3.89} & \textbf{9.14 $\pm$ 2.08}  

    \label{tab:voxres}
    \end{tabular}
    \end{center}
    \vspace{-0.4cm}
\end{table*} 

\endgroup

\subsection{Main Results}

The results in all three downstream tasks, for all tested configurations, are displayed in Tab.~\ref{tab:mixres} for the case of mixture input, and Tab.~\ref{tab:voxres} for the case of vocal input. In both tables, the first row corresponds to the COLA~\cite{saeed21} setup, the second to MSCOL~\cite{garoufis23}, the third to MERT~\cite{li23} and the fourth to the COLA-like artist identity-informed sampling scheme presented in~\cite{alonso22}. Lines 5-7 correspond to the label-agnostic CVSM variants, whereas the performance of the label-informed CVSM variant is presented on the last line. Both the best results across label-informed and label-agnostic methods are typeset in bold. We observe that in both mixture and vocal cases, the label-agnostic CVSM variants outperform the COLA~\cite{saeed21} and MERT~\cite{li23} baselines in the majority of downstream tasks (the exception being artist similarity for CVSM-A), while performing comparably to MSCOL~\cite{garoufis23} on musical mixtures, and better on isolated vocals. Similarly, the proposed model pre-trained on the association between mixtures and isolated vocals from the same artist (CVSM-ART) was competitive to its respective label-informed baseline (COLA-ART~\cite{alonso22} for the case of musical mixtures (see Tab.~\ref{tab:mixres}), while outperforming it for isolated vocals (Tab.~\ref{tab:voxres}); we note that, in accordance to the literature~\cite{alonso22,meehan25}, the availability of artist labels during pre-training leads to improvement in the examined downstream tasks.

Upon comparison of the two tables, we observe that for the classification tasks (gender identification and artist identification), the best performance reached by any CVSM variant on musical mixtures (see Tab.~\ref{tab:mixres}) is comparable to that achieved with isolated vocals (see Tab.~\ref{tab:voxres}). This indicates that CVSM is suitable for end-to-end application in vocal understanding tasks, without necessitating a vocal isolation pre-processing step. On the other hand, this does not hold entirely true in the artist similarity task, denoting that isolating the vocals as a pre-processing step is still necessary for successful retrieval applications.

Delving into more detail in the model performance across the evaluated downstream tasks, we first examine the results for the case of full musical mixtures, which are reported in Tab.~\ref{tab:mixres}. We first note that the incorporation of vocal excerpts in label-agnostic training pipelines leads to improved performance in both gender and artist-related tasks. Indeed, CVSM outperforms both COLA~\cite{saeed21} and MERT~\cite{li23}, which have been trained through full mixtures, while the performance yielded by MSCOL~\cite{garoufis23}, which has been trained through mixture-vocal pairs, is comparable to some CVSM variants. On the contrary, both label-informed models yield comparable performance in all tasks, suggesting that the variation of contrastive pairs obtained through artist identity sampling is sufficient. Regarding the different tasks, the advantage of the label-informed models (CVSM-ART and COLA-ART) is clearer in gender identification, where all label-agnostic CVSM variants perform to similar levels to MSCOL~\cite{garoufis23}. However, examination of the results in artist identification and retrieval reveal a more complex picture; while solely incorporating artificial mixtures of vocals and instrumental accompaniment (CVSM-A) does lead to slightly better performance in artist identification compared to MSCOL~\cite{garoufis23}, it comes at the cost of a more fragmented latent space, as implied by the higher EER and MNR metrics in artist similarity. Both hybrid real-artificial pre-training (CVSM-AH) and in-domain finetuning (CVSM-AF) help in resolving this issue, leading to better metrics in both identification and similarity than MSCOL. Among these two methods, CVSM-AH appears more effective; in fact, in artist identification, its results approach the ones achieved by the label-informed approaches, despite not accessing identity labels during pretraining. The performance yielded by CVSM-AH in comparison to both MSCOL~\cite{garoufis23} and CVSM-A implies in contrast to the majority of augmentations devised in~\cite{spijkervet21}, increasing the probability of generating artificial pairs as an augmentation does not guarantee better downstream performance. Finally, the advantage that the label-informed models hold in artist similarity is relatively significant, amounting to a more than 4\% overall decrease in both EER and MNR metrics compared to their label-agnostic counterparts.

The overall picture changes slightly in the case of isolated vocal input, as we can deduce from the results presented in Tab.~\ref{tab:voxres}. Here, i) the availability of vocals during pre-training, as well as ii) the application of augmentations that disentangle their properties, emerge as crucial factors of model performance. In more detail, CVSM-ART outperforms COLA-ART~\cite{alonso22} in all examined downstream tasks, highlighting the importance of incorporating isolated vocal excerpts into pre-training; this trend is maintained on the label-agnostic models, with CVSM variants and MSCOL~\cite{garoufis23} scoring higher than both COLA~\cite{saeed21} and MERT~\cite{li23}. In addition, in both artist identification and artist similarity tasks, all CVSM variants yield better results compared to MSCOL~\cite{garoufis23}, indicating that the proposed scheme of creating artificial musical mixtures, and correlating them with vocal excerpts, succeeds in isolating vocal-related attributes. Contrary to the case of musical input mixtures, the performance levels between CVSM-A and the two more sophisticated variants are relatively similar, which we ascribe to the exposure of CVSM-A in clean vocals (as opposed to non-realistic artificial mixtures) during pre-training. Finally, the performance of CVSM-ART in the gender identification (which requires a coarsely structured latent space) and artist similarity tasks suggests that in terms of latent space structure, availability of artist labels does play a crucial role; on the other hand, as we also observed in the complete mixture case, the label-agnostic CVSM variants achieve performance close to CVSM-ART in the artist identification task.

\subsection{Quantitiative Analysis} 

The results presented previously were obtained, assuming the availability of song clips of sufficient length (30 sec). However, in practice (i.e. for real-time applications), successfully infering information from shorter audio segments is also necessary. To this end, we experimented with varying the \textit{clip length} used for aggregating the per-excerpt predictions into a single one. In addition to the above, we investigated the performance of the embeddings CVSM generates under a \textit{low-resource setup}, as is common practice in the literature~\cite{spijkervet21,xiao22}. For this purpose, we re-trained linear classifiers on top of the frozen encoders in the task of artist identification, using the same artist splits as in the previous experiments, but with a decreased portion of data available for training and validation. Note that throughout these experiments, we mainly want to compare contrastively pre-trained models that integrate vocal information in their training scheme; as such, we exclude MERT~\cite{li23} and COLA-ART~\cite{alonso22} from those, keeping COLA~\cite{saeed21} as a baseline reference. We also omit results for the CVSM-AF variant, since it exhibited a similar trend to CVSM-AH.

\begin{figure}[t!]
\begin{minipage}{0.49\linewidth}
        \centering
        \centerline{\includegraphics[width=4.3cm]{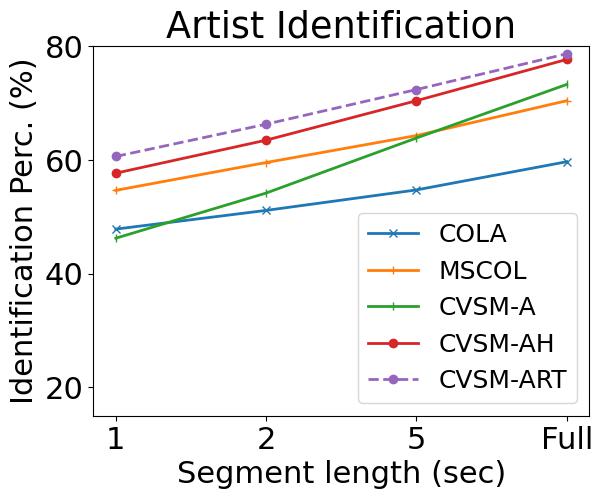}}
    \end{minipage}
    \begin{minipage}{0.49\linewidth}
        \centering
        \centerline{\includegraphics[width=4.3cm]{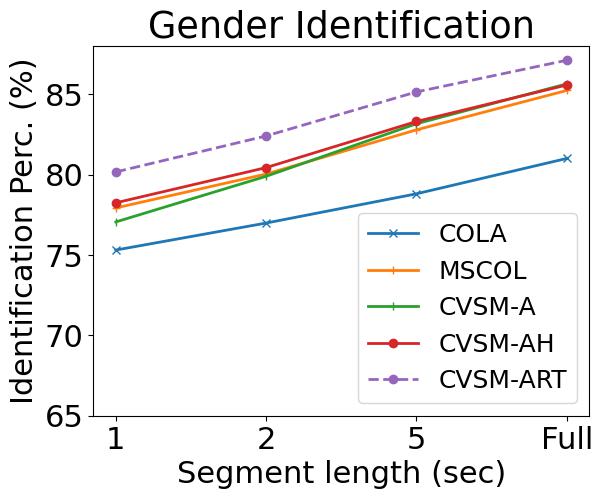}}
\end{minipage}
    \caption{\textcolor{black}{Performance on the tasks of artist identification (left) and \textcolor{black}{gender identification (right)}, depending on the length of input context available to the network (in sec).}}
    \vspace{-0.35cm}
    \label{fig:evol}
\end{figure}

The performance of CVSM embeddings of different origins, aggregated through various clip durations is visualized in Fig.~\ref{fig:evol}, for the case of mixture inputs and the tasks of artist (left) and gender (right) identification. In both cases, aggregation of the per-excerpt scores for longer inputs leads to higher performance in the artist identification task, as it stabilizes network predictions. However, we observe that the performance of all variants does not rise consistently in relation to the excerpt length. In more detail, in the task of artist identification, the embeddings obtained from CVSM-A record lower identification performance than the COLA embeddings obtained without any explicit vocal excerpts, when subjected to single excerpts (1~sec. duration). On the other hand, aggregating embeddings from a successively larger amount of excerpts yields significantly better performance than COLA, even improving over MSCOL and approaching CVSM-AH. We hypothesize that this is related to the informative, yet ``noisier'' latent space of CVSM-A, since it has not encountered any real music mixtures during pre-training. A similar trend can be deduced for the task of gender identification; while the CVSM-A embeddings, obtained prior to in-domain finetuning, record the best scores among label-agnostic models after prediction aggregation, their relative performance to both MSCOL and CVSM-AH drops when reducing the duration of available audio, reaching a negative difference of -1.5\% for 1 sec. excerpts. Finally, the performance advantage of the embeddings obtained via artist-informed pretraining, CVSM-ART, is consistent throughout all clip lengths; in fact, in the case of artist identification, its gap to label-agnostic models is larger for smaller-length inputs, reaching approximately 3\% for 1 sec excerpts.

\begin{figure}[t!]
\begin{minipage}{0.49\linewidth}
        \centering
        \centerline{\includegraphics[width=4.3cm]{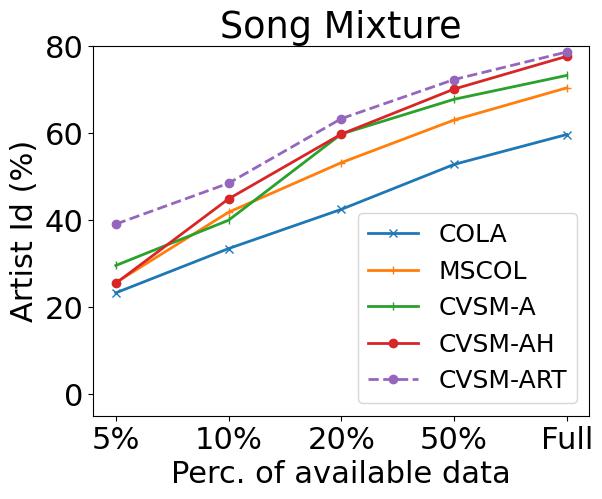}}
    \end{minipage}
    \begin{minipage}{0.49\linewidth}
        \centering
        \centerline{\includegraphics[width=4.3cm]{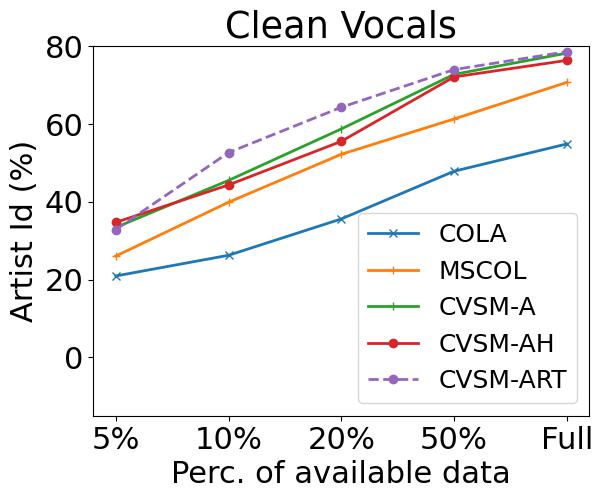}} 
\end{minipage}
    \caption{\textcolor{black}{Performance of the obtained frozen embeddings on the task of artist identification, subject to a reduced data regime, when using the full mixture (left) or the vocal excerpts (right) as network input.}}
    \vspace{-0.35cm}
    \label{fig:dataevol}
\end{figure}

 For the low-resource experiments, the results are visualized in Fig.~\ref{fig:dataevol}, for both mixture input (left) or isolated vocals (right). In both cases, we observe that while for an adequate amount of available data the gap between CVSM-ART and the label-agnostic CVSM variants remains close, CVSM-ART performs significantly better under low-resource settings; this result is aligned with the higher effectiveness of CVSM-ART in the artist similarity task. Interestingly, among label-agnostic models, the performance gap between MSCOL and CVSM tends to decrease for smaller data percentages, suggesting the dilution of the latent space with artificial examples requires a higher amount of labeled examples to unlock its performance advantage. A possible explanation for this could lie in the lack of musicality in the generated examples (since they are formed by random vocal-accompaniment superimposition), which could be potentially resolved through selection of appropriate multi-tracks for mixture generation~\cite{ciranni24,riou24}.

\begin{figure}
\begin{minipage}{0.49\linewidth}
        \centering
        \centerline{\includegraphics[width=4.9cm]{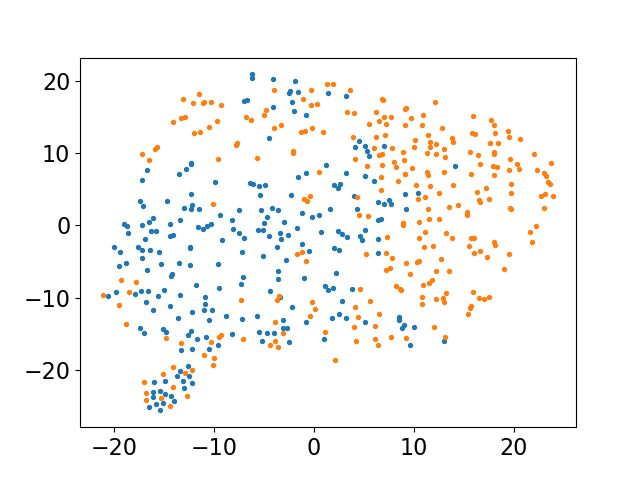}}
    \end{minipage}
    \begin{minipage}{0.49\linewidth}
        \centering
        \centerline{\includegraphics[width=4.9cm]{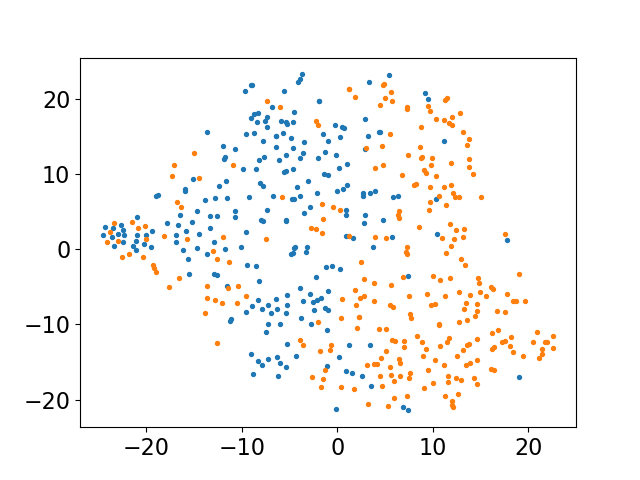}}
\end{minipage}
\begin{minipage}{0.49\linewidth}
        \centering
        \centerline{\includegraphics[width=4.9cm]{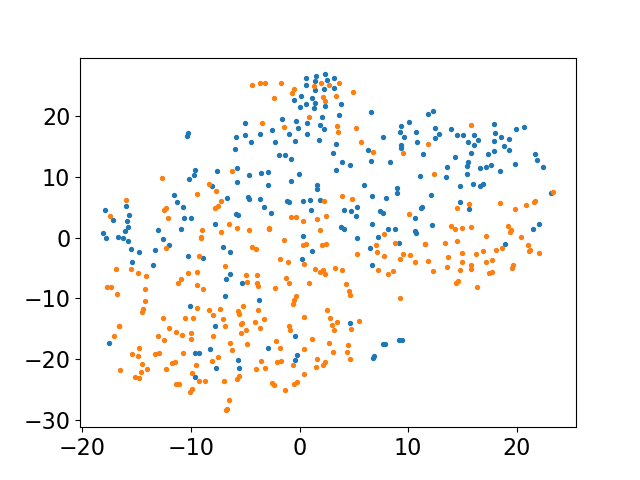}}
    \end{minipage}
    \begin{minipage}{0.49\linewidth}
        \centering
        \centerline{\includegraphics[width=4.9cm]{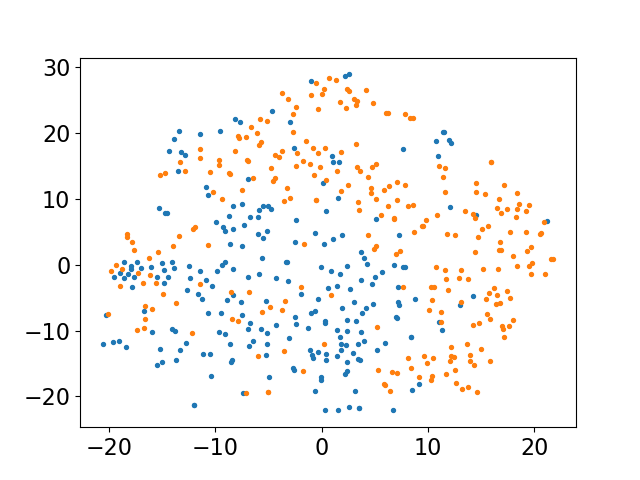}}
\end{minipage}
\begin{minipage}{0.49\linewidth}
        \centering
        \centerline{\includegraphics[width=4.9cm]{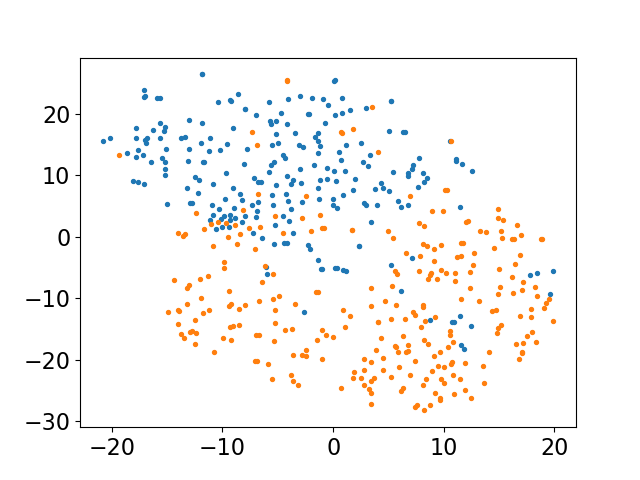}}
    \end{minipage}
    \begin{minipage}{0.49\linewidth}
        \centering
        \centerline{\includegraphics[width=4.9cm]{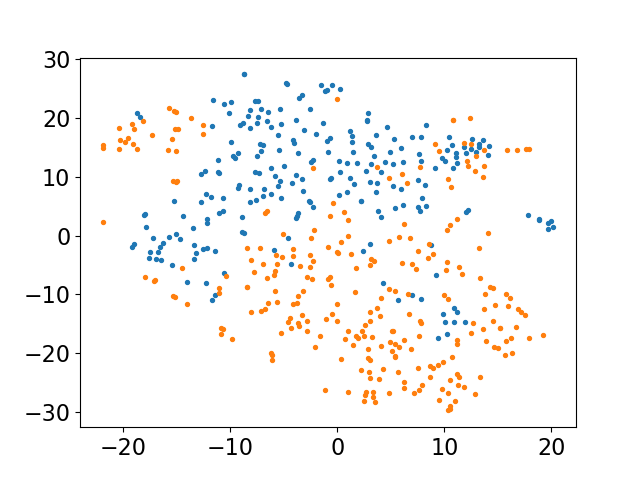}}
\end{minipage}
    \caption{T-SNE projections of clip-wise average embeddings from various models; blue dots correspond to male singers, orange to female. The top row plots correspond to CVSM-A (top left) and CVSM-AH (top right) variants, the middle row to the label-agnostic baselines COLA (middle left) and MSCOL (middle right), and the bottom row to the label-informed models, COLA-ART (bottom left) and CVSM-ART (bottom right).}
    \label{fig:tsnes_gender}
    \vspace{-0.4cm}
\end{figure}

\begin{figure}
\begin{minipage}{0.49\linewidth}
        \centering
        \centerline{\includegraphics[width=4.9cm]{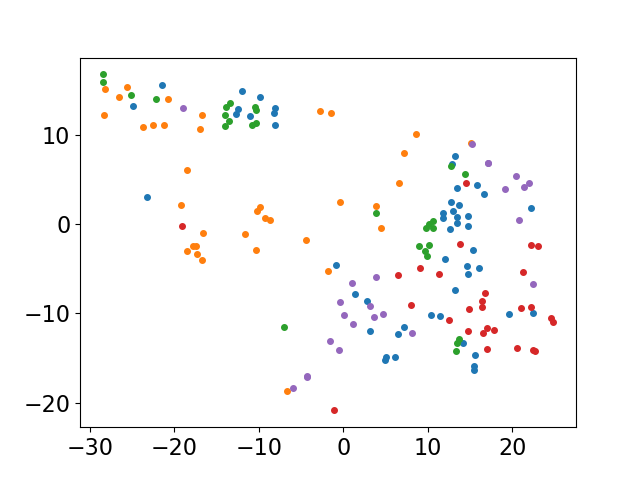}}
    \end{minipage}
    \begin{minipage}{0.49\linewidth}
        \centering
        \centerline{\includegraphics[width=4.9cm]{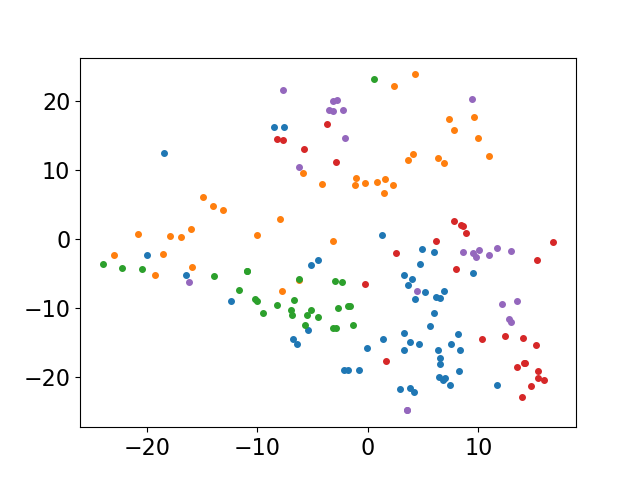}}
\end{minipage}
\begin{minipage}{0.49\linewidth}
        \centering
        \centerline{\includegraphics[width=4.9cm]{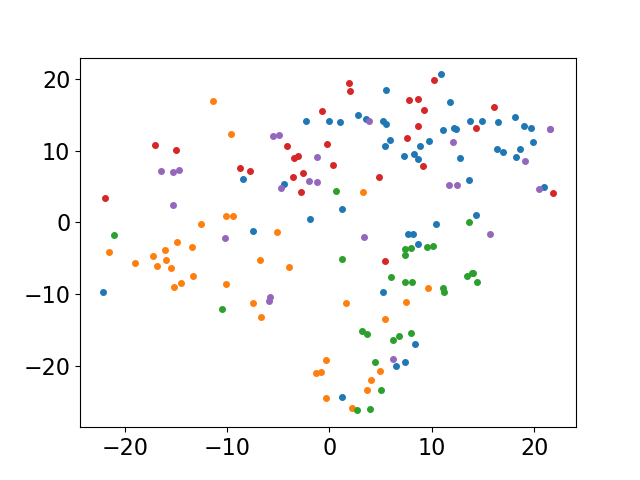}}
    \end{minipage}
    \begin{minipage}{0.49\linewidth}
        \centering
        \centerline{\includegraphics[width=4.9cm]{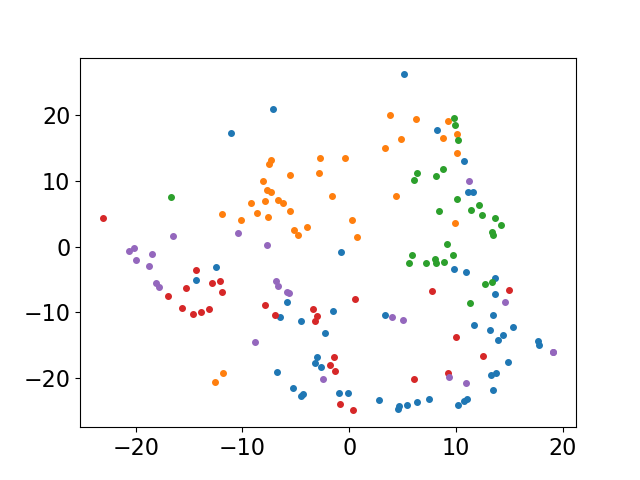}}
\end{minipage}
\begin{minipage}{0.49\linewidth}
        \centering
        \centerline{\includegraphics[width=4.9cm]{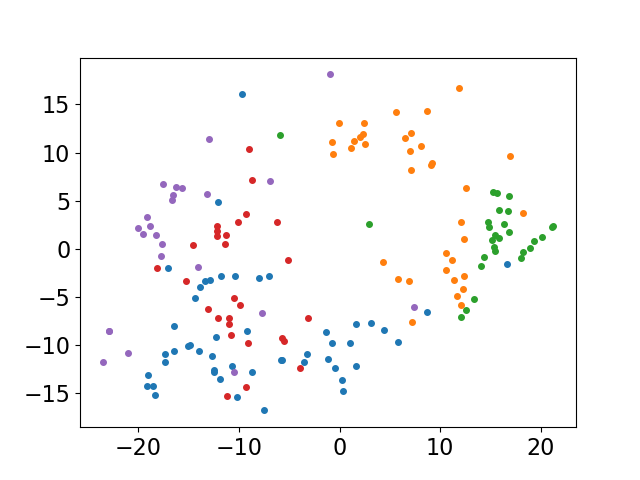}}
    \end{minipage}
    \begin{minipage}{0.49\linewidth}
        \centering
        \centerline{\includegraphics[width=4.9cm]{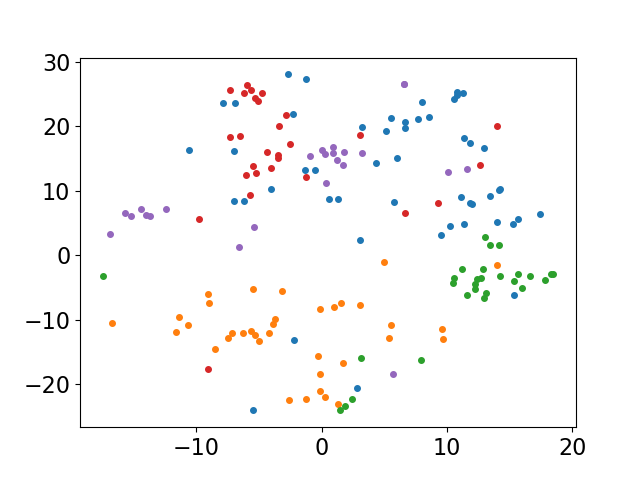}}
\end{minipage}
    \caption{T-SNE projections of clip-wise average embeddings from various models; dots of the same color correspond to the same artist. The top row plots correspond to CVSM-A (top left) and CVSM-AH (top right) variants, the middle row to the label-agnostic baselines COLA (middle left) and MSCOL (middle right), and the bottom row to the label-informed models, COLA-ART (bottom left) and CVSM-ART (bottom right).}
    \vspace{-0.4cm}
    \label{fig:tsnes_art}
\end{figure}

\begin{figure*}[t!]
\begin{minipage}{0.33\linewidth}
        \centering
        \centerline{\includegraphics[width=5.3cm]{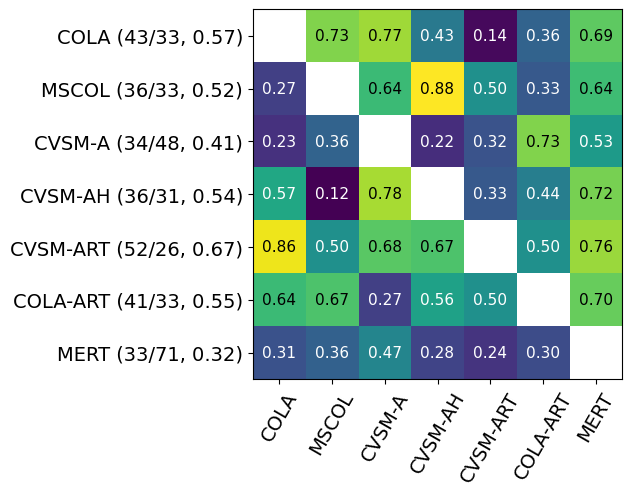}}
    \end{minipage}
    \begin{minipage}{0.33\linewidth}
        \centering
        \centerline{\includegraphics[width=5.3cm]{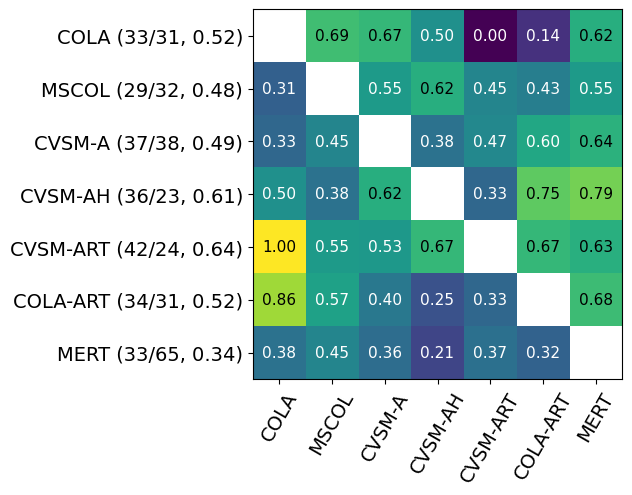}} 
\end{minipage}
    \begin{minipage}{0.33\linewidth}
        \centering
        \centerline{\includegraphics[width=5.3cm]{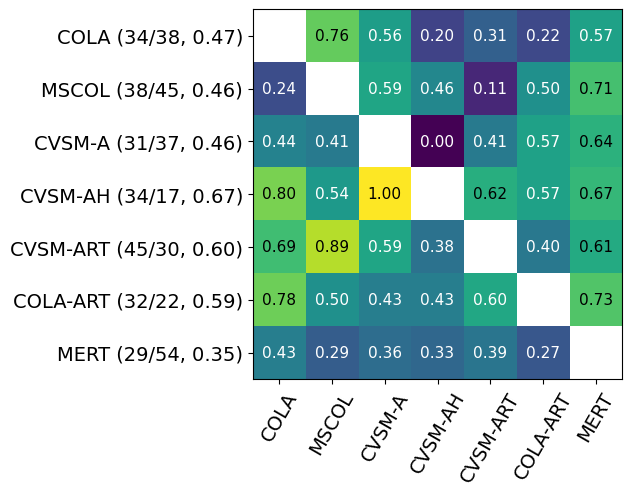}} 
\end{minipage}
    \caption{Pairwise comparison results between the models evaluated on the listening test, regarding overall similarity between musical mixtures (left), vocal similarity between overall mixtures (center) and similarity between isolated vocals (right); values in each cell denote the percentage of instances the model in the respective row was chosen over the one in the respective column, the overall scores for each model (in terms of both wins/losses and winrate percentage) are displayed in parentheses at the left of each row.}
    \vspace{-0.35cm}
    \label{fig:pairwiseres}
\end{figure*}

\subsection{Latent Space Visualizations}

To visualize the extent CVSM manages to effectively create a latent space with disentangled singer attributes, we plot the T-SNE~\cite{vandermaaten08} projections of the clipwise average embeddings of complete musical mixtures (probed from the encoder's output layer), for a randomly sampled subset of the Music4All validation split, conditioned on either the \textit{gender of the singer}, or the \textit{artist identity}. To measure the quality of the generated clusters, we measure the per-cluster average silhouette score~\cite{rousseeuw87} as well as the per-cluster average ratio between the mean intra-cluster and the mean inter-cluster distances, averaged across 5 T-SNE runs. Again, with a similar rationale to before, we do not present these visualizations for MERT~\cite{li23} and CVSM-AF; COLA-ART~\cite{alonso22} is included, for a qualitative comparison to COLA-ART.

In Fig.~\ref{fig:tsnes_gender}, we display the results for the case of gender labels, with clips corresponding to male singers displayed in blue dots, while those of female singers colored in orange. The top row corresponds to T-SNE plots for label-agnostic CVSM variants, either solely using artificial mixtures (CVSM-A, top left) or combining artificial with real mixtures (CVSM-AH, top right) for pre-training; the middle row contains embeddings for the COLA-trained (middle left) and MSCOL-trained (middle right) baselines, while the bottom row displays the embedding distribution for the label-informed COLA-ART (bottom left) and CVSM-ART (bottom right) models. We observe that there is high overlap between male and female voices for both COLA and MSCOL baselines (average silhouette scores and distance ratios around 0.10), suggesting that external factors, such as a song's tempo or instrumentation, are influencing the structure of the learned latent space. On the other hand, for the top two plots, the male- and female- voiced clips occupy slightly more segregated areas in the T-SNE plot. This distinction is equally discernible for CVSM-AH (top right), indicating that despite exposure to contrastive pairs sampled from the same song, the learned latent space retains vocal-specific properties. However, none of the label-agnostic models create separate clusters between male and female voices, which we attribute to the absence of explicit supervisory signals. This distinction is a bit more visible for the label-informed method, resulting in slightly more structured latent spaces, with comparable performance (silhouette scores and distance ratios between 0.2-0.3 across runs -- reaching a distance ratio of 0.290 for COLA-ART). In addition, despite a relatively clear (with minimal overlap) border between the two clusters, no separate areas are formed in the T-SNE plot. Sub-clusters may also appear, such as the female-majority area in the top left of the CVSM-ART plot.

Similarly, in Fig.~\ref{fig:tsnes_art}, we present the embedding T-SNE projections with respect to the artist identity of the input audio (differently colored dots corresponding to different artist identities), with the same model correspondence as in Fig.~\ref{fig:tsnes_gender}. Among the various models, we observe that again, the models trained on artist labels achieve the best separation among different artist classes (silhouette score of 0.128 and inter-intra ratio of 0.468 for CVSM-ART; silhouette score of 0.112 and inter-intra ratio of 0.478 for COLA-ART), with some of those occupying distinct subspaces in the T-SNE plot. These results are consistent with those of the artist similarity tasks, wherein models utilizing artist identity labels during pre-training yielded much better EER and MNR scores. Among label-agnostic models, the performance achieved is relatively similar, with the largest separation achieved in the case of CVSM-AH (mean silhouette and intra-inter scores of 0.040 and 0.334 respectively), which are significantly lower than the ones yielded by label-informed models. For the rest of the label-agnostic models, we generally observe higher overlap between different artist classes.

\section{Subjective Evaluation}

The complete results of the pairwise comparisons conducted during the listening tests are portrayed in Fig.~\ref{fig:pairwiseres}. The two subfigures on the left and the center correspond to the perceived similarity between complete musical mixtures (the left one regarding the overall similarity, and the center one on the vocal similarity), while the right one on the respective results on isolated vocals. Each row and column on the heatmap corresponds to a different model, with the percentage displayed in each cell denoting the winrate of the model in its row against the model in its column; the overall score for each model (both in terms of wins/losses and overall win percentage) is displayed in parentheses at the left of each row. From the results, we observe the following:

\begin{itemize}
\item In general, models with artist identity information during pre-training appear to outperform their identity-agnostic counterparts in detecting similarity in musical mixtures. This is apparent from the positive overall winrates for both CVSM-ART and COLA-ART models; in the case of CVSM-ART, the achieved scores (52/26 overall similarity and 42/24 vocal similarity, with all pairwise comparisons involving it either favoring it or being tied) supersede random chance at the $p=0.05$ statistical significance level.
\item The label-agnostic models, while mostly lagging behind the label-informed ones, display slight deviations depending on whether the target similarity concerns the overall properties of the musical piece, or just the vocals. In more detail, CVSM-A performs the worse among all contrastively pre-trained models in modeling overall similarity, whereas surprisingly, COLA~\cite{saeed21} yields the best results amongst all label-agnostic models. On the other hand, CVSM-AH outperforms all other label-agnostic models on vocal similarity, with the rest of the contrastive models (COLA~\cite{saeed21}, MSCOL~\cite{garoufis23}, and CVSM-A) yielding effectively similar results. The performance of COLA~\cite{saeed21} in overall similarity, compared to models that incorporate vocals into their training pipeline, implies that vocal similarity modeling ``sacrifices'' perceptually important instrumental information on the instrumental. Conversely, the relative improvement achieved by both CVSM-A and CVSM-AH in vocal similarity modeling, compared to the overall one, suggests that the proposed pre-training scheme with artificial mixtures partially succeeds in creating a latent space with perceptually important vocal information; in the case of CVSM-AH, the quality of the latent space in this aspect approaches the one of CVSM-ART. We note, however, that with the exception of CVSM-AH in the case of vocal similarity (statistically significant at the $p=0.05$ level), none of the other overall scores are statistically significantly different to random chance.
\item In the case where the query and the retrieved pieces consist of isolated vocals, the results of the user study largely coincide with those obtained from evaluating vocal similarity in complete musical mixtures. In particular, CVSM-ART among label-informed models and CVSM-AH among label-agnostic ones exhibit the best performance, at a $p < 0.05$ statistical significance level compared to random chance; other contrastive models display performance close to random chance.
\item Finally, we observe that under both examined scenarios, all contrastive models achieve superior performance to MERT~\cite{li23}, indicating that contrastive pre-training paradigms might be more suitable for similarity modeling and retrieval applications.
\end{itemize}

\begin{figure}[t!]
\begin{minipage}{0.49\linewidth}
        \centering
        \centerline{\includegraphics[width=4.5cm]{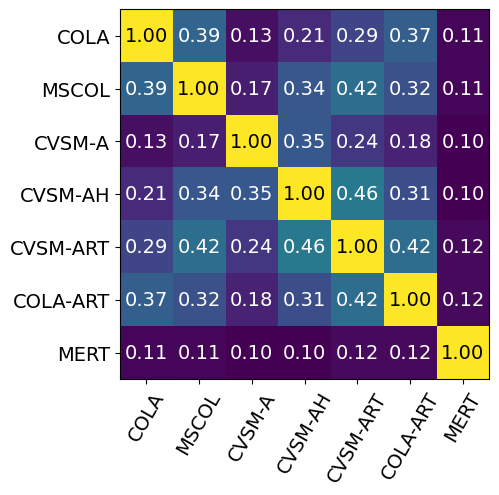}}
    \end{minipage}
    \begin{minipage}{0.49\linewidth}
        \centering
        \centerline{\includegraphics[width=4.5cm]{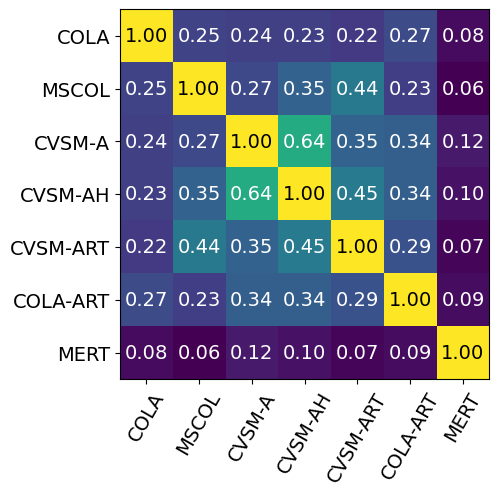}}
\end{minipage}
    \caption{Pairwise comparison of the recommendations given by the evaluated models, among the queries presented in the subjective test, for mixture input (left) and vocal input (right); higher values correspond to higher recommendation similarity.}
    \vspace{-0.45cm}
    \label{fig:modelsim}
\end{figure}

As we have mentioned, the subjective test uses primarily sample triplets where the compared models provide different recommendations, mostly excluding the instances where the same recommendation is retrieved by different models. Thus, as an additional similarity measure of the latent spaces between the various models, we calculated the percentage of instances among the query set where the same recommendation is obtained for each model pair. These similarities are depicted in the heatmaps presented in Fig.~\ref{fig:modelsim}, where each row and each column corresponds to a different model; the left heatmap presents the similarities for mixture input, the right heatmap on isolated vocals. We observe that under both settings, the highest similarity to the CVSM-ART model, which demonstrated consistent performance in both objective and subjective evaluation setups, is achieved by the CVSM-AH model, followed by MSCOL~\cite{garoufis23}. This suggests that enriching an identity label-free pre-training dataset with artificial mixtures of vocals and accompaniment may contribute to approximating the latent space of a label-informed model. Interestingly, while CVSM-A displays a relatively small portion of recommendations that are similar to the rest of the models in the case of mixture input, presumably due to the lack of exposure in real musical mixtures, its similarity with CVSM-AH in the case of vocal input is particularly high, with the two models retrieving the same recommendation in 64\% of the input queries. Finally, we observe that in both cases, MERT~\cite{li23} yields the lower number of shared recommendations across all models; we attribute this to the different training scheme followed in~\cite{li23}, not involving contrastive losses, as well as the larger excerpt length used for pre-training~\cite{li23}.

\section{Conclusions}

In this work, we explored the applicability of contrastive learning into learning representations of musical audio with respect to attributes of the singing voice, and presented CVSM, a framework that learns such representations by maximizing the similarity between musical mixtures and vocal excerpts. We devised both a \textit{label-informed} pre-training scheme, which leverages artist labels during contrastive pair sampling, and a \textit{label-agnostic} protocol based on generating artificial mixtures through superimposing isolated vocals and randomized instrumental accompaniment. Our results, validated both through linear probing on downstream tasks encapsulating aspects of vocal similarity and a user study, suggest that the proposed method is effective in conveying vocal similarity, performing at least comparably to the respective (pending on artist label availability during pre-training) baselines. We also note that while the label-informed scheme exhibited more consistent performance across both downstream testing and the user study, a hybrid, label-agnostic pre-training scheme with a combination of real and artificial musical mixtures performed competitively to it both in the artist identification task and in perceived vocal similarity. This work contributes in paving the way towards modeling musical similarity with respect to particular sources, or other fine-grained musical attributes.

Since the obtained results indicate that availability of artist labels during pre-training leads to more consistent downstream performance and similarity modeling, additional work should be carried out towards bridging the gap between label-informed and label-agnostic protocols, by incorporating an identity estimation step during contrastive sampling. Moreover, integration of the artificial mixture creation pipeline into label-informed pre-training should be further investigated, since our preliminary experiments (not reported here) yielded a performance drop in downstream testing. Finally, it would be interesting to adapt the proposed strategy in other music information retrieval tasks involving frame-wise information from vocals, such as automatic lyrics transcription~\cite{mesaros10} and vocal fundamental frequency estimation~\cite{cuesta20}. 

\section{Acknowledgments}

The authors would like to thank Panagiotis P. Filntisis for his very useful feedback on the content of the paper.

\bibliography{refs}
\bibliographystyle{IEEEtran}

\end{document}